# Solar Energy, Wind, Eruptions, Magnetic Fields, Neutrinos, and Planetary Material: Surface Evidence of an Iron-Rich Solar Interior and a Neutron-Rich Solar Core

O. Manuel and Y. Singh (University of Missouri-Rolla)

Events at the solar surface likely reflect those in the interior of the Sun. In addition to the well-known solar neutrino puzzle [1], however, the model of a hydrogen-filled Sun [2] does not explain solar magnetic fields and eruptions [3], primordial variations in elements and isotopes linked across planetary distances [4], nor the annual outpouring of $3 \times 10^{43}$ H atoms [5] from the solar surface. Outer regions of the planetary system are rich in light elements (H, He, C). The inner region is rich in heavier elements (Fe, S, Si). Isotope ratios of yet heavier elements (Te, Xe, Ba) in meteorite inclusions of C are unlike those in FeS inclusions [6]. These differences are repeated in the atmospheres of Jupiter [7] and Mars [6]. These regions never mixed: They are outer and inner regions of the supernova [8] that produced the solar system ≈ 5 Gy ago [9]. Terrestrial planets accreted heterogeneously [10]. The iron cores formed first and then acted as accretion sites for silicates that formed further from the Sun. The iron-rich Sun [11] accreted on the collapsed SN core. Its magnetic fields are deep-seated remnants from the core and/or a rotating, superfluid, superconductor made by Bose-Einstein condensation of material around the core [12]. Neutron-emission from the core [13] triggers a series of reactions that produce solar luminosity, the carrier gas for solar mass separation, and the outflow of H from the solar surface:

1. Neutron emission from the solar core
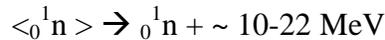
$$<{}_0^1n> \rightarrow {}_0^1n + \sim 10\text{-}22 \text{ MeV}$$
2. Neutron decay or capture
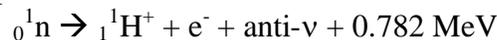
$${}_0^1n \rightarrow {}_1^1H^+ + e^- + \text{anti-}\nu + 0.782 \text{ MeV}$$
3. Fusion and upward migration of $H^+$
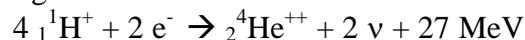
$$4\,{}_1^1H^+ + 2\,e^- \rightarrow {}_2^4He^{++} + 2\,\nu + 27 \text{ MeV}$$
4. Escape of excess $H^+$ in the solar wind
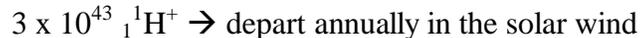
$$3 \times 10^{43}\,{}_1^1H^+ \rightarrow \text{depart annually in the solar wind}$$

Detection of inverse β−decay induced by low-energy (E < 0.782 MeV), antineutrinos coming from the Sun (e.g., the production of 87-day $^{35}$S in the Homestake Mine [14] or in underground salt deposits by capture of anti-neutrinos on $^{35}$Cl) would confirm the second step of this process.

**References**: [1] T. Kirsten, *Rev. Mod. Phys.* **71**, 1213 (1999).
[2] A. Dar and G. Shaviv, *Ap. J.* **468**, 935 (1996).
[3] L. Lanzerotti, *Space Ref. com.* (14 Nov. 2003) See comments on solar magnetic fields and eruptions: http://www.spaceref.com/news/viewpr.html?pid=13022.
[4] J. T. Lee et al., *Comments Astrophys.* **18**, 335-345 (1997).
[5] J. Geiss and P. Boschler, in *The Sun in Time*, p. 98 (eds: C. P. Sonnett et al., Univ. Arizona Press, Tucson, AZ, 1991).